\documentclass[showpacs,prl,twocolumn,aps,superscriptaddress]{revtex4}
\usepackage{amsmath,amssymb,graphicx}
\begin{document}
\title{Strong Mobility in Weakly Disordered Systems}
\author{E.~Ben-Naim}
\affiliation{Theoretical Division and Center for Nonlinear
Studies, Los Alamos National Laboratory, Los Alamos, New Mexico
87545}
\author{P.~L.~Krapivsky}
\affiliation{Department of Physics,
Boston University, Boston, Massachusetts 02215}
\begin{abstract}
We study transport of interacting particles in weakly disordered
media. Our one-dimensional system includes (i) disorder: the hopping
rate governing the movement of a particle between two neighboring
lattice sites is inhomogeneous, and (ii) hard core interaction: the
maximum occupancy at each site is one particle.  We find that over a
substantial regime, the root-mean-square displacement of a particle,
$\sigma$, grows super-diffusively with time $t$, $\sigma\sim
(\epsilon\,t)^{2/3}$, where $\epsilon$ is the disorder
strength. Without disorder the particle displacement is sub-diffusive,
$\sigma\sim t^{1/4}$, and therefore disorder dramatically enhances
particle mobility.  We explain this effect using scaling arguments,
and verify the theoretical predictions through numerical
simulations. Also, the simulations show that disorder generally leads
to stronger mobility.
\end{abstract}
\maketitle 

Disorder, inhomogeneities, and impurities are ubiquitous in physical
systems and are widely used to control properties of matter.  Some of
the most fascinating phenomena in contemporary physics including
localization \cite{pwa,dt,aalr}, glassiness \cite{ks,gp}, slow
relaxation \cite{fh}, and frustration \cite{apr} are unique
consequences of disorder.

While the effects of disorder on noninteracting particles are
well-understood, the consequences of disorder on interacting, strongly
correlated particles remain an open question \cite{kssmf,lr}.  In a
quantum system, an isolated particle is localized by disorder, but
localization can be destroyed when there are two interacting particles
\cite{dls,ps}. Hence, disorder and particle interactions compete. We
investigate this interplay between disorder and interaction in a
classical system where inhomogeneities are known to trap particles and
drastically decrease their mobility, and find that, as in the quantum
case, disorder has opposite effects on noninteracting and on
interacting particles. Our main result is that disorder speeds up the
motion of interacting particles whereas disorder slows down the
movement of noninteracting particles.

\begin{figure}[t]
\includegraphics[width=0.4\textwidth]{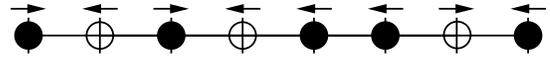}
\caption{Illustration of the disordered interacting particle
system. The arrows indicate the bias at each site, the circles
indicate vacant sites, and the bullets indicate occupied sites.}
\label{fig-lattice}
\end{figure}

We generalize the standard exclusion process \cite{bd,gms} to study
the interplay between disorder and interactions
\cite{tb,jk,ed,mb}. Our system is an unbounded one-dimensional lattice
whose sites may be either occupied by a single particle or
vacant. Initially, the lattice is populated at random by identical
particles with concentration $c$. Each particle may hop from an
occupied site into a neighboring vacant site and this diffusion
process is governed by the following rates: $p_+(i)$ is the hopping
rate from site $i$ to site $i+1$, and similarly, $p_-(i)$ is the
hopping rate from site $i$ to site $i-1$. While the total hopping rate
is uniform, and is set to one without loss of generality,
$p_-(i)+p_+(i)=1$, the lattice is inhomogeneous. At every site there
is, with equal probabilities, a bias to the right, $p_+=1/2+\epsilon$,
or a bias to the left, $p_+=1/2-\epsilon$, as illustrated in figure
\ref{fig-lattice}.  The parameter $0\leq \epsilon\leq 1/2$ is the
disorder strength.  Note that the disorder is quenched, uncorrelated,
and uniform in strength. Moreover, since every lattice site
accommodates a single particle, the particles interact via hard core
repulsion \cite{tml}. Our problem generalizes two well-known
processes: single-file diffusion
\cite{teh,dgl,ap,vkk,ra,wbl,cdl,bhmst} with interaction but no
disorder corresponds to the limit $\epsilon\to 0$ , and Sinai
diffusion \cite{ygs,aog,hk,flm} with disorder but no interaction
corresponds to the limit $c\to 0$.

Our focus is transport in this disordered, interacting particle,
system. Since there is no overall bias in either direction, on
average, the displacement of a particle with respect to its initial
position, $x$, does not change with time, $\langle x\rangle=0$.  We
ask the most elementary question: how does the root-mean-square
displacement, $\sigma$, defined by $\sigma^2=\langle x^2\rangle$,
evolve? We address this question via a scaling analysis of weakly
disordered systems, $\epsilon\ll 1$, and numerical simulations with
general disorder strengths.

\noindent{\bf Early Times.} When disorder is weak, $\epsilon \ll 1$,
there is an initial period during which particles do not ``feel'' the
disorder and hence move at random, $p_+=p_-=1/2$. In this early
regime, disorder is irrelevant and the behavior is dominated by
particle interactions.  Without disorder, the hard core repulsion
causes a dramatic change in mobility: whereas an isolated particle
moves diffusively, $\sigma \sim t^{1/2}$, the root-mean-square
displacement of an interacting particle grows sub-diffusively with
time \cite{teh,dgl,ap,vkk,ra}
\begin{equation}
\label{early}
\sigma\sim t^{1/4}.
\end{equation}
Therefore, the movement of a particle is severely hindered by the
presence of other particles. We illustrate this remarkable collective
behavior for extremely dense systems \cite{dgl} where there are large
clusters of occupied sites that are separated by isolated vacancies.
Particles move by exchanging their position with neighboring
vacancies. Furthermore, the sparse vacancies can be regarded as
non-interacting \cite{dgl}. A particle that, up to time $t$, exchanges
position with a total of $N=N_++N_-$ vacancies of which $N_+$ were
initially located to its right and $N_-$ were initially located to its
left, has the displacement $x=N_+-N_-$. First, since the vacancies are
randomly distributed in the initial configuration, the excess of
vacancies in one direction follows from the law of large numbers,
$|N_+-N_-|\sim N^{1/2}$, and consequently, $x\sim N^{1/2}$. Second,
vacancies that were initially located at a distance on the order of
the diffusive length scale $t^{1/2}$ from a particle may exchange
position with it. Therefore, $N\sim (1-c)\,t^{1/2}$ and combining this
scaling law with $x\sim N^{1/2}$ yields \eqref{early}. Although this
scaling argument applies to densely packed systems, the behavior
\eqref{early} holds for arbitrary concentrations
\cite{teh,dgl,ap,vkk}. We also comment that this suppressed diffusion
is a direct consequence of the hard core interactions.

\noindent{\bf Intermediate Times.} Eventually, the disorder becomes
relevant, and the biased hopping rates do affect the particle
displacement. Although there is no global bias in the hopping rates,
there certainly are local biases, as illustrated in figure
\ref{fig-lattice} where sites with negative bias are in the majority.
We expect that, at least at intermediate times scales, or
equivalently, intermediate length scales, these local biases lead to
directed motion \cite{bg,mm,sr}.

To quantify how such local biases affect particle mobility, we
consider a particle that visits $\sigma$ distinct sites of which $n_+$
have a positive bias and $n_-$ have a negative bias with
$\sigma=n_++n_-$. Since the disorder is uncorrelated, the difference
between the number of positive and negative sites, $\Delta=|n_+-n_-|$,
grows diffusively with the total number of visited sites, $\Delta\sim
\sigma^{1/2}$.  The excess of sites biased in one direction leads to a
drift in this preferred direction with the small velocity
$v\sim\epsilon\, \Delta/\sigma$ or \hbox{$v\sim
\epsilon\,\sigma^{-1/2}$}.  Furthermore, the ballistic length scale
$x\sim v\,t$ gives an estimate for the displacement, $x \sim
(\epsilon\,t)\,\sigma^{-1/2}$. Since the displacement must be of the
same order as the total number of sites visited, $x\sim \sigma$, we
have
\begin{equation}
\label{self-consistent}
\sigma\sim \epsilon\,t\,\sigma^{-1/2}.
\end{equation}
We thus arrive at our main result: the displacement becomes
super-diffusive because of the disorder,
\begin{equation}
\label{intermediate}
\sigma\sim (\epsilon\,t)^{2/3}.
\end{equation}
Of course, this length scale ultimately exceeds the suppressed
diffusion length scale \eqref{early}.  We conclude that the
inhomogeneous hopping rates generate a stochastic local velocity
field, and as a result, there are local drifts that significantly
enhance the mobility of the particles.

\noindent{\bf Late Times.}  To understand the behavior at late times,
we recall that the displacement of a non-interacting particle in a
random disorder is logarithmically slow \cite{ygs,aog,hk}
\begin{equation}
\label{late}
x\sim \epsilon^{-2}(\ln t)^2.
\end{equation}
At sufficiently large length scales, the random disorder generates a
potential well that confines the particle. The depth of this potential
well is the sum of all the biases in a given range, $U(x)=\sum_{i=1}^x
\left(p_+(i)-p_-(i)\right)$, and therefore, the depth of the well
grows diffusively with distance, $U\sim \epsilon \sqrt{x}$. This
stochastic well constitutes a barrier that the particle must overcome,
and since the time to escape out of this barrier grows exponentially
with the depth of the well, $t\sim \exp(U)\sim
\exp(\epsilon\sqrt{x})$, the displacement is logarithmic as in
\eqref{late}.

We argue that the slow mobility \eqref{late} also characterizes the
asymptotic late time behavior of interacting particles in
disorder. First, the confining potential well remains the same even
when there are multiple particles. Second, the probability that a
given particle escapes the well is exponentially small, and therefore,
only mildly affected by the presence of other particles. We envision a
scenario where particles are stuck in a local minimum of the potential
and escape the barrier one at a time. Such an escape process is
dominated by the same exponential escape time that characterizes an
isolated, non-interacting particle. We conclude that at late times,
interacting particles in a random disorder also follow the logarithmic
displacement law \eqref{late}.  Particle interactions become
irrelevant and the behavior is governed by disorder alone.

\begin{figure}[t]
\includegraphics[width=0.4\textwidth]{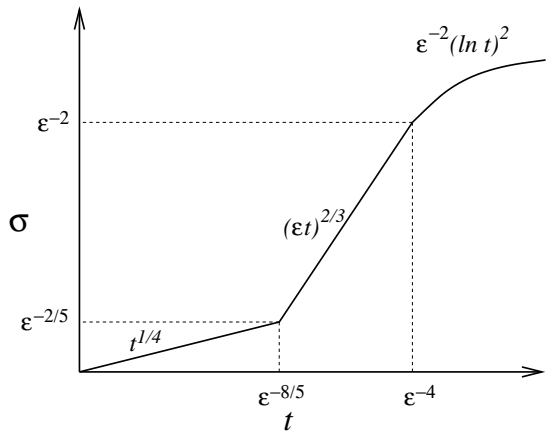}
\caption{The three regimes of behavior \eqref{three}.  The
displacement $\sigma$ is plotted versus time $t$ using a double
logarithmic scale.}
\label{fig-three}
\end{figure}

\noindent{\bf The Three Time Regimes.} By combining the early
\eqref{early}, intermediate \eqref{intermediate}, and late
\eqref{late} time behaviors, we conclude that the mobility of a given
particle exhibits three distinct regimes of behavior as follows (see
also figure \ref{fig-three}),
\begin{equation}
\label{three}
\sigma \sim
\begin{cases}
t^{1/4}              & t\ll \epsilon^{-8/5},\\
(\epsilon\, t)^{2/3} & \epsilon^{-8/5}\ll t\ll \epsilon^{-4},\\
 \epsilon^{-2}(\ln t)^2    &\epsilon^{-4}\ll t.
\end{cases}
\end{equation}
The time and length scales that characterize the crossover points can
be obtained by matching the two corresponding behaviors.  The
transition from the early regime into the intermediate regime occurs
at time $t\sim \epsilon^{-8/5}$ and length $\sigma\sim
\epsilon^{-2/5}$, while the transition from the intermediate domain
into the late domain occurs at time $t\sim \epsilon^{-4}$ \cite{log}
and length $\sigma\sim \epsilon^{-2}$, as shown in figure
\ref{fig-three}.

Let us recap the three regimes of behavior. At the early stages,
particle interactions dominate over disorder, and the motion of
particles is sub-diffusive due to the hard core repulsion.  In the
intermediate regime, disorder and interactions are both relevant. The
particles stream following the stochastic local velocity field and the
result is a strong, super-diffusive transport. At late times, disorder
dominates and interactions become irrelevant. Particles are trapped by
a stochastic potential well and the displacement is logarithmically
slow because the escape time is exponentially large.

As further support of the scaling behavior above, we can show that the
stochastic potential well plays no role in the intermediate
regime. Clearly, since the overall hopping rate equals one, the time
scale characterizing the movement between neighboring sites is also of
order one. The time to escape out of a well grows exponentially with
the depth of the well, $t\sim \exp(U)$, but this time scale becomes
appreciable only when the depth of the potential well is large,
$\epsilon\sqrt{x}\gg 1$, or equivalently, when the displacement
becomes sufficiently large, $x\gg \epsilon^{-2}$. Indeed, this length
scale is realized only at the late time regime, as shown in figure
\ref{fig-three}. Therefore, trapping is negligible throughout the
intermediate regime.

Let us now consider the effect of disorder on a non-interacting
particle.  In the absence of disorder, $\epsilon=0$, the particle
displacement is unhindered and thus, purely diffusive, $\sigma\sim
t^{1/2}$. In weak disorder, $\epsilon\ll 1$, an isolated particle
undergoes ordinary diffusion at early times, but is later slowed down
considerably according to \eqref{late}. Hence, there are two distinct
regimes of behavior when interactions are absent
\begin{equation}
\label{two}
\sigma \sim
\begin{cases}
t^{1/2}&t\ll\epsilon^{-4},\\
\epsilon^{-2}(\ln t)^2& \epsilon^{-4}\ll t.
\end{cases}
\end{equation}
We note that the crossover time scale matches the upper time scale in
\eqref{three}.  Thus, in the absence of particle interactions,
disorder slows the particles down.

Surprisingly, disorder has the opposite effect on an interacting
particle system.  Due to disorder, particles undergo fast,
super-diffusive motion as in \eqref{intermediate} over a substantial
time range and this effect becomes stronger as the disorder weakens
because the crossover time and length scales are divergent.  According
to \eqref{three}, the displacement in a given disorder eventually
overtakes the displacement in a weaker disorder. This non-monotonic
dependence on the disorder strength is another nontrivial consequence
of the competition between disorder and interactions.  We conclude
that in the presence of particle interactions, disorder speeds the
particles up.

\begin{figure}[t]
\includegraphics[width=0.38\textwidth]{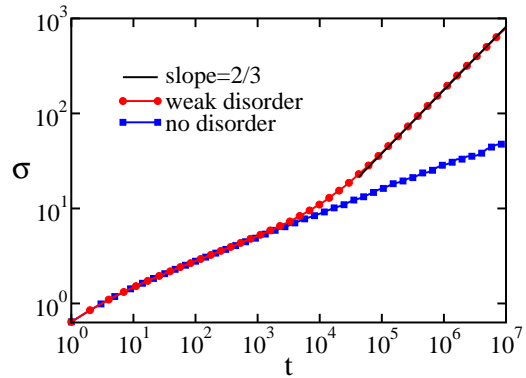}
\caption{The early and intermediate behaviors for weak disorder,
$\epsilon=10^{-2}$ (bullets) and no disorder, $\epsilon=0$
(squares). Shown is the displacement $\sigma$ versus time $t$
(bullets), as well as a reference line with slope $2/3$.}
\label{fig-weak}
\end{figure}

\noindent{\bf Numerical Simulations.} We performed extensive Monte
Carlo simulations to test the scaling predictions. The simulations are
a straightforward implementation of the transport process. Initially,
identical particles randomly occupy the sites of a one-dimensional lattice
of size $L$ with periodic boundary conditions, and the initial
concentration equals $c$. Each lattice site has a bias in the positive
or the negative direction as $p_+=1/2+\epsilon$ or $p_-=1/2-\epsilon$
with equal probabilities. The dynamics are asynchronous. In an
elementary step, a randomly chosen particle hops to the right with
probability $p_+$ or to the left with probability $p_-$, and this hop
is successful only if the neighboring site is vacant. Subsequently,
time is augmented by the inverse number of particles, $t\to
t+1/N$. This elementary step is repeated indefinitely. We present
results of simulations with $L=4\times 10^5$ and $c=1/2$.

We verified the super-diffusive behavior \eqref{intermediate} using a
weak disorder (figure \ref{fig-weak}). Even though the super-diffusive
behavior is an intermediate asymptotic, the duration of this regime
grows rapidly as the disorder weakens. We performed a few additional
tests: (i) we checked that the displacement $\sigma$ is a function of
the scaled time variable $\epsilon\, t$ rather than $t$ at
intermediate times by using different disorders, (ii) we verified that
the concentration does not play an important role using $c=1/4$ and
$c=3/4$, and (iii) we used a different type of disorder with $p_+$
drawn from a flat distribution in the range
$[1/2-\epsilon:1/2+\epsilon]$ and obtained qualitatively similar
results.

\begin{figure}[t]
\includegraphics[width=0.36\textwidth]{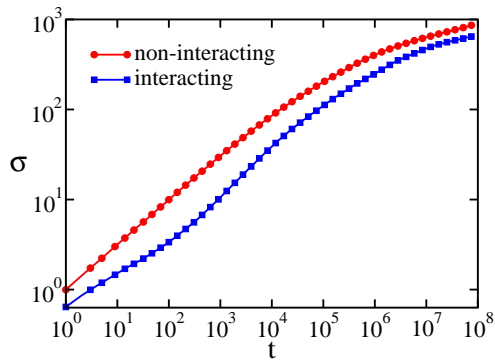}
\caption{The late time behavior for $\epsilon=10^{-1}$. Shown is the
displacement $\sigma$ versus time $t$ for non-interacting
particles (squares) and for interacting particles (bullets).}
\label{fig-late}
\end{figure}

\begin{figure}[t]
\vspace{.1in}
\includegraphics[width=0.36\textwidth]{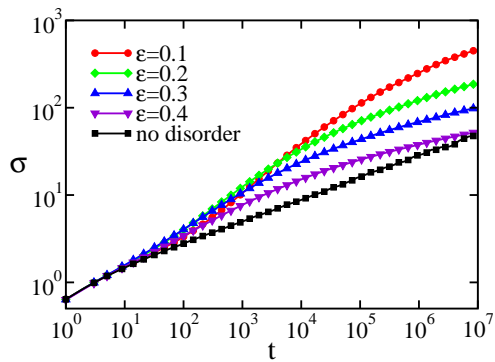}
\caption{The behavior at moderate disorders. Shown is the displacement
$\sigma$ versus time $t$ without disorder ($\epsilon=0$, squares) and
with moderate disorder values of $\epsilon=0.1$ (bullets), $0.2$
(diamonds), $0.3$ (down-triangle), and $0.4$ (up-triangle).}
\label{fig-moderate}
\end{figure}

To test the behavior at late times, we also simulated a
non-interacting particle system by ignoring the site occupancy
restriction.  These simulations show that after an extremely long
transient period, the displacements of interacting particles and
non-interacting particles nearly match (figure \ref{fig-late}),
thereby confirming that hard core interactions become irrelevant
asymptotically, and that the behavior is governed by disorder alone.

Our scaling analysis tacitly assumes that disorder is small.  A
comparison of the behaviors with moderate disorders and with no
disorder shows that, irrespective of the disorder strength, mobility
is always strengthened by disorder (figure \ref{fig-moderate}). Thus,
mobility enhancement is a general effect that does not require weak
disorder.

Figure \ref{fig-moderate} also shows that the displacement in a homogeneous
system eventually catches up with the displacement in a strongly
inhomogeneous system. Indeed, the sub-diffusive behavior \eqref{early}
that characterizes a uniform system eventually exceeds the logarithmic
displacement \eqref{late} in a disordered system. However, the
crossover time $t\sim \epsilon^{-8}$ is astronomical at weak disorders
and in practice, disorder always generates a stronger
mobility. Indeed, the crossover time is very large even
at moderate and strong disorders (figure \ref{fig-moderate}).

In conclusion, we studied how disorder affects transport in an
interacting particle system. We found that whereas disorder slows down
non-interacting particles, disorder speeds up interacting particles.
Therefore, there is an intricate interplay between interaction and
disorder.

Disorder provides an effective mechanism for controlling transport
properties because weak disorders result in strong mobilities. This
effect can be tested experimentally in colloidal \cite{wbl} or
biological \cite{cdl,bhmst} channels, where the slow transport
\eqref{early} was realized.

\acknowledgments We thank Nigel Goldenfeld, Sidney Redner, and Stuart
Trugman for useful discussions.  We are grateful for financial support
from DOE grant DE-AC52-06NA25396, NSF grants CHE-0532969 and
CCF-0829541.


\begin{thebibliography}{99}
\bibitem{log} Taking logarithmic corrections into account, the second
      transition time is $t\sim \epsilon^{-4}[\ln (1/\epsilon)]^3$.
\end{thebibliography}
\end{document}